\documentclass[prb,aps,final,floatfix]{revtex4}
\input epsf
\begin{document}
\def\beq{\begin{equation}}
\def\eeq{\end{equation}}
\def\etr{\varepsilon_{\rm tr}}
\def\erstar{\varepsilon_r^*}
\def\Gam{\Gamma(\erstar,J_r)}

\title{Statistical evaporation of rotating clusters. IV. Alignment
effects in the dissociation of nonspherical clusters}

\author{P. Parneix}
\affiliation{Laboratoire de Photophysique Mol\'eculaire\footnote{laboratoire 
associ\'e \`a l'universit\'e de Paris-Sud}, CNRS, F\'ed\'eration
de recherche Lumi\`ere Mati\`ere, B\^at. 210,
Universit\'e Paris-Sud, F91405 Orsay Cedex, France}
\author{F. Calvo}
\affiliation{Laboratoire de Physique Quantique, IRSAMC, Universit\'e Paul
Sabatier, 118 Route de Narbonne, F31062 Toulouse Cedex, France}

\begin{abstract}
Unimolecular evaporation in rotating, non-spherical atomic clusters is
investigated using Phase Space Theory in its orbiting transition state
version. The distributions of the total kinetic energy release
$\etr$ and the rotational angular momentum $J_r$ are calculated for
oblate top and prolate top main products with an arbitrary degree of
deformation.
The orientation of the angular momentum of the product cluster with
respect to the cluster symmetry axis has also been obtained.
This statistical approach is tested in the case of the small
8-atom Lennard-Jones cluster, for which comparison with extensive
molecular dynamics simulations is presented. The role of
the cluster shape has been systematically studied for larger, model
clusters in the harmonic approximation for the vibrational densities
of states. We find that the type of deformation (prolate vs. oblate)
plays little role on the distributions and averages of $\etr$ and
$J_r$ except at low initial angular momentum. However, alignment
effects between the product angular momentum and the symmetry axis
are found to be significant, and maximum at some degree of oblateness.
The effects of deformation on the rotational
cooling and heating effects are also illustrated.
\end{abstract}
\maketitle

\section{Introduction}

Statistical theories of unimolecular dissociation\cite{forst,smith}
provide invaluable information about binding energies, lifetimes, and
more generally the thermodynamical state of atomic and molecular
clusters.\cite{jarroldrev} Besides their common use in helping to
interpret experiments,\cite{brech,vogel1} their quantitative relevance
has been demonstrated in numerical simulations\cite{wa,ph} with great
accuracy. Among them, the phase space theory (PST) of
Nikitin,\cite{nikitin} Light and coworkers,\cite{light}
Klots,\cite{klots} and Chesnavich and Bowers\cite{cb} stands out as
the most successful\cite{wa,ph} in describing evaporation of weakly
bound clusters for which the loose transition state hypothesis is
satisfied. Within the PST formalism, statistical dissociation is
characterized by the vibrational densities of states (VDOS) of the
parent and product clusters, as well as the rotational density (RDOS)
of the products. This latter quantity depends on the interaction
between the fragments, and more importantly on their shapes,\cite{cb}
at least through their number of rotational degrees of freedom.\cite{klots}

Large clusters of simple materials, such as rare gases or alkali
metals, are usually found to be rather spherical.
Small aggregates can nevertheless exhibit significant deformations
at the scale of a few tens of atoms or molecules. The shape of
covalent clusters could be inferred from the ion mobility measurements
performed by the Jarrold group.\cite{jarroldsi,jarroldc,jarroldtin}
In silicon clusters, a prolate-oblate transition was evidenced around 20
atoms.\cite{jarroldsi} Small carbon clusters show a competition
between linear chains, rings and planar shapes.\cite{jarroldc} Larger
carbon clusters can be arbitrarily prolate under the nanotube
geometry. Deformed metal clusters have been investigated theoretically
using various approaches,\cite{naprb} especially in concern with the
fission problem.\cite{naher}

Accounting for deviations from the spherical shape may be crucial
for calculating precisely the number of rovibrational states of
small molecules, and the role of nonsphericity on the statistical
observables relevant to unimolecular dissociation has been
discussed previously. Chesnavich
and Bowers\cite{cb,cbrev} quantified the error introduced in the
rotational DOS when approximating symmetric rigid rotors by spherical
tops, for small molecules or radicals with moderate deformations.
Berblinger and Schlier\cite{berblinger}
used a Monte Carlo integration method to compute the density of
quantum states of the oblate molecules H$_3^+$ or HD$_2^+$ using
appropriate densities or rotational states.

Peslherbe and Hase considered evaporation in small aluminium
clusters from the point of view of molecular dynamics (MD) simulations
and PST.\cite{ph} In this work, deformation effects were not fully
accounted for, because the centrifugal barrier was neglected except in
the spherical top approximation for the fragments. However it is known
that centrifugal forces, and more generally rotation may affect
cluster properties quite significantly.\cite{jellinek,wales,calvomc}

In our previous work, an extensive study of evaporation in rotating
atomic and molecular clusters was carried out,\cite{pap1,pap2,pap3}
paying a special attention to the final angular momentum and kinetic
energy released. By carefully calculating the vibrational and
rotational DOS's pertaining to various cases of spherical clusters
emitting atoms,\cite{pap1,pap2} linear or spherical
molecules,\cite{pap3} we could extend the conclusion of Weerasinghe
and Amar\cite{wa} that PST is quantitative in reproducing the exact
numerical results. However some clear deviations could be seen in the
statistical treatment of evaporation when applied to the smaller
cluster LJ$_8$, where the main product LJ$_7$ was approximated as a
spherical top. This poorly satisfactory result was attributed to the
non-spherical character of LJ$_7$, known to be oblate in its most
stable energy structure. 

Our first goal in this paper is to incorporate the effects of
deformations in the PST treatment for rotating clusters. As a first
application, atomic dissociation in LJ$_8$ will be considered as an
example. More generally, we will focus on large clusters simplified as
molten, homogeneous droplets, in order to investigate the role of
nonsphericity on the observables relevant to cluster
dissociation. Following previous effort,\cite{wa,ph,pap3,pap4} it is hoped
that cluster evaporation could be used as a probe of the shape, and
possibly the phase changes, in the main product. The dynamical
simulation of fragmentation in asymetric molecules such as those
aforementioned involve can be very challenging from a computational
point of view. Therefore it is important to incorporate and quantify the
effects of deformation in the rate theories used to interpret
gas-phase experiments on clusters.

In the next Section, we summarize the theoretical results of phase
space theory in the approximation where the main product after atom
dissociation can be modelled as a prolate or oblate top. Concise
expressions for the rotational densities of states, KER and final
angular momentum are proposed. We also compute the distribution of the
relative orientation between the symmetry axis and the angular
momentum in the product. This property, along with its average value,
provide an estimation of how the rotational energy is distributed in
the cluster. We illustrate the specific case of LJ$_8$ and larger LJ
clusters in Sec.~\ref{sec:res}. In particular, the roles of size,
total energy and angular momentum on the final properties are
discussed. Finally, we summarize and give some concluding remarks in
Sec.~\ref{sec:ccl}.

\section{Methods}
\label{sec:met}

\subsection{Phase Space Theory}

The rotational density of states (RDOS) is a key function within the
framework of phase space theory.\cite{nikitin,light,klots,cb}
This quantity depends on the kinetic energy released after
dissociation, $\etr$, sum of
the translational and rotational parts. It also
depends on the total angular momentum $J$. It is
obtained by integration of the rotational sum of states
$\Gamma(\erstar,J_r)$, which quantifies the number of rotational
states available at a given
angular momentum $J_r$ of the product cluster, and a kinetic energy
lower than $\erstar$, its limiting value at the centrifugal barrier.\cite{cb}

In PST the two fragments are treated as rigid bodies and the expressions
of $\Gamma$ only depend on their corresponding symmetries. Our present
interest concerns the evaporation from nonspherical atomic clusters,
and we have chosen to deal with symmetric top deformations, prolate
or oblate, but with an arbitrary degree of nonsphericity.
The rotational constants are denoted as $A$ and $B$ for a prolate top
cluster ($\lambda=A-B\geq 0$), and as $B$ and $C$ for an oblate top
cluster ($\lambda=B-C\geq 0$). The rotational energy
$\varepsilon_r$ is expressed by $\varepsilon_r=BJ_r^2
\pm \lambda K^2$, where $J_r$ and $K$ are the internal angular
momenta, with + (resp. $-$) for the prolate top (resp. oblate top) case.

One motivation of the present work is to investigate the possible
correlations between the cylindrical symmetry axis of the main
product cluster and the angular momentum vector after fragmentation.
Therefore the most general statistical quantity we need to consider is
the probability for a dissociation event to occur with prescribed
values of $K$, $J_r$, and $\etr$:
\begin{equation}
P(K,J_r,\etr;E,J)dKdJ_rd\etr \propto \omega(E-E_0-\etr) dKdJ_rd\etr
\int_{\cal L} dL,
\label{eq:pkjr}
\end{equation}
where $\omega$ is the vibrational density of states of the product
cluster, $E_0$ the energy difference between the parent and product,
and ${\cal L}$ the available values of orbital momentum $L$ at
fixed $J_r$, $\etr$, and $K$. 
The RDOS results from summing the above equation over all $(K,J_r)$
values compatible with energy and angular momentum conservation. 
For this we first need to integrate over $K$ for rotational energies
no larger than $\erstar$.\cite{cb} The expressions of these
sums of states are given by
\begin{equation}
\Gamma = \left\{ \begin{array}{ll}
2J_r = \Gamma_{\rm S}(\erstar,J_r) & ~~\mbox{if~} \erstar \geq AJ_r^2 \\
\displaystyle 2\left[ \frac{\erstar - BJ_r^2}{\lambda} \right]^{1/2} =
\Gamma_{\rm P}(\erstar,J_r) & ~~\mbox{if~} AJ_r^2\geq \erstar \geq BJ_r^2
\end{array}\right.
\label{eq:sumpa}
\end{equation}
for the prolate top+atom case, and by
\begin{equation}
\Gamma = \left\{ \begin{array}{ll}
2J_r = \Gamma_{\rm S}(\erstar,J_r) & ~~\mbox{if~} \erstar \geq BJ_r^2 \\
\displaystyle 2J_r - 2\left[ \frac{BJ_r^2 - \erstar}{\lambda} \right]^{1/2} =
\Gamma_{\rm S}(\erstar,J_r)-\Gamma_{\rm O}(\erstar,J_r) & ~~\mbox{if~}
BJ_r^2\geq \erstar \geq CJ_r^2 \end{array}\right.
\label{eq:sumoa}
\end{equation}
for the oblate top+atom case.

\begin{figure}[htb]
\vbox to 11cm{
\includegraphics{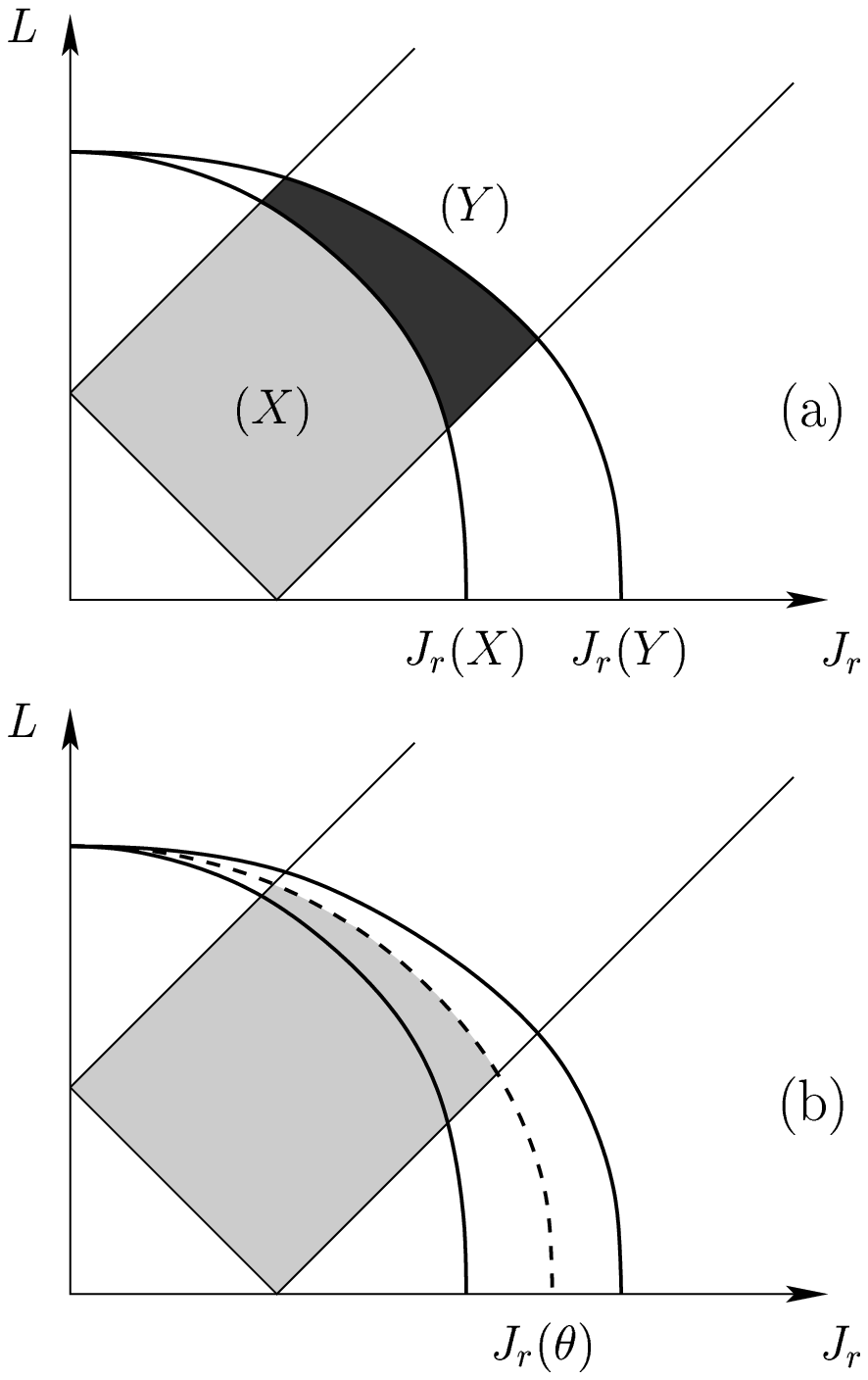}
\vfill}
\caption{Range of integration in the ($L,J_r$)
plane for atom+symetric top fragments. (a) Complete integration range,
$X>Y$ being the two rotational constants. (b) Restricted integration
range when the orientation of $J_r$ with respect to the symmetry
axis is restricted to have a fixed angle $\theta$.}
\label{fig:schema}
\end{figure}

To obtain the rotational densities of states at given $\etr$ and $J$,
the sums of rotational states must be integrated in the whole range
${\cal C}$ of available values of $J_r$ and $L$, accounting for
conservation of total energy and angular momentum.
The general schematic integration plot is represented in
Fig.~\ref{fig:schema}(a). Here $X$ and $Y$ are the two rotational
constants of the product, such that $X>Y$, equal to $A$ and $B$
(resp. $B$ and $C$) in the prolate top (resp. oblate top) case.
The barrier height corresponding to orbital
momentum $L$ is denoted as $\varepsilon^\dagger(L)$ in what follows.
The procedure used to calculate the RDOS at finite angular momentum
is basically the same as used previously by Chesnavich and
Bowers,\cite{cb} and by ourselves.\cite{pap1,pap3}

The intersections $J_r(X)$ and $J_r(Y)$ have the respective values
$(\etr/X)^{1/2}$ and $(\etr/Y)^{1/2}$. We introduce the limits
$L_X^-$ and $L_X^+$ (resp. $L_Y^-$ and $L_Y^+$) as the intersections
of $\etr = XJ_r^2 + \varepsilon^\dagger(L)$ (resp.
$\etr = YJ_r^2 + \varepsilon^\dagger(L)$) with $L=|J-J_r|$ (for
$L_X^-$ and $L_Y^-$) and with $L=J+J_r$ (for $L_X^+$ and
$L_Y^+$). At low $\etr$, corresponding to $\etr \leq
\varepsilon^\dagger(J)$, $L_X^+$ and $L_Y^+$ are also given from
the intersections with $L=|J-J_r|$.

Moreover we must remind that when $J$ is not zero, the KER
$\etr$ has a minimum value $\etr^{\min}$ where integration starts.
This value actually depends on the smallest rotational constants,
namely $Y$. $\etr^{\min}$ is such that the curve $\erstar=YJ_r^2$ is
tangent to the line $L=J-J_r$.

The rotational sums of states, Eqn.~(\ref{eq:sumpa}) and
(\ref{eq:sumoa}), can be gathered into
$\Gamma(\erstar,J_r)=\Gamma_{\rm s}(\erstar,J_r) \pm
\Gamma_{\rm t}(\erstar,J_r)$, where $\Gamma_{\rm s}=2J_r$ is the sphere+atom
part, and $\Gamma_{\rm t}$ the non-spherical, symmetric top part.
In the definition of $\Gamma$, a plus sign is used for the prolate
top, and a minus sign for the oblate top. According to this simple
partition, the RDOS can then be expressed as a sum of two terms:
\begin{eqnarray}
\Gamma(\etr,J) &=& \displaystyle
\int\!\!\!\!\int_{\cal C} \Gamma(\erstar,J_r) dJ_r dL \nonumber \\
&=& \Gamma_{\rm s}(\etr,J) \pm \Gamma_{\rm t}(\etr,J).
\label{eq:rdostot}
\end{eqnarray}

The contribution of the spherical top part to the rotational density
of states has been calculated previously,\cite{pap1} and reads
\begin{equation}
\Gamma_{\rm s}(\etr,J) = \Gamma^-_{\rm s} + \Theta(\etr - B_sJ^2)
\Gamma^+_{\rm s},
\label{eq:rdoss}
\end{equation}
where $\Theta$ is the Heaviside function, and where $B_s$ is the
rotational constant equal to $A$ (prolate top) or $C$ (oblate top).
The two components $\Gamma^-_{\rm s}$ and $\Gamma^+_{\rm s}$ of
$\Gamma_{\rm s}$ are given by
\begin{equation}
\Gamma^-_{\rm s}(\etr,J) = \int_{L_{B_s}^-}^{L_{B_s}^+} \left[
\frac{\etr-\varepsilon^\dagger(L)}{B_s} - (J-L)^2\right] dL,
\label{eq:gasm}
\end{equation}
and
\begin{equation}
\Gamma^+_{\rm s}(\etr,J) = 2J(L_{B_s}^-)^2,
\label{eq:gasp}
\end{equation}
respectively.

Similarly, the non-spherical RDOS $\Gamma_{\rm t}(\etr,J)$ can be
expressed as a sum of two terms, namely $\Gamma_{\rm t} =
\Gamma^-_{\rm t} + \Theta(\etr - YJ^2) \Gamma^+_{\rm t}$. These
two terms cannot be fully explicited in general, due to the unknown (or
complex) dependence of $\varepsilon^\dagger$ or $\erstar$ with $L$.
However, integral forms are readily available from
\begin{equation}
\Gamma^-_{\rm t}(\etr,J) = \int_{L_Y^-}^{L_Y^+} dL \int_{\rm
max(|J-L|, \sqrt{\erstar/X})}^{(\erstar/Y)^{1/2}} 2 \left|
\frac{\erstar - BJ_r^2}{\lambda}\right|^{1/2} dJ_r
\label{eq:gatm}
\end{equation}
and
\begin{equation}
\Gamma^+_{\rm t}(\etr,J) = \int_{L_X^-\Theta(\etr - XJ^2)}^{L_Y^+}
dL \int_{\rm max(|J-L|, \sqrt{\erstar/X})}^{J+L} 2 \left|
\frac{\erstar - BJ_r^2}{\lambda}\right|^{1/2} dJ_r.
\label{eq:gatp}
\end{equation}

The integrals over $J_r$ are solved to yield
\begin{equation}
\Gamma^-_{\rm t}(\etr,J)=\frac{1}{\sqrt{B\lambda}} \int_{L_Y^-}^{L_Y^+} dL
\erstar \left[ \Psi\left( \sqrt{\frac{B}{Y}}\right) - \Psi \left(
\max \left( |J-L|\sqrt{\frac{B}{\erstar}},\sqrt{\frac{B}{X}}\right)
\right) \right],
\label{eq:gamtm2}
\end{equation}
\begin{equation}
\Gamma^+_{\rm t}(\etr,J)=\frac{1}{\sqrt{B\lambda}}
\int_{L_X^-\Theta(\etr-XJ^2)}^{L_Y^+} dL
\erstar \left[ \Psi\left( (J+L)\sqrt{\frac{B}{\erstar}}\right) - \Psi \left(
\max \left( |J-L|\sqrt{\frac{B}{\erstar}},\sqrt{\frac{B}{X}}\right)
\right) \right],
\label{eq:gamtp2}
\end{equation}
with the function $\Psi$ given by

\begin{equation}
\Psi(x)=x\sqrt{1-x^2} + \sin^{-1} x
\end{equation}
for prolate top fragments, and by
\begin{equation}
\Psi(x)=x\sqrt{x^2-1} -\ln (x+\sqrt{x^2-1})
\end{equation}
for oblate top fragments.

\subsection{Kinetic energy release distribution}

From the density of rotational states, the kinetic energy release distribution
can be easily calculated from the usual relationship
\begin{eqnarray} P(\etr;E,J) \propto
\int_{\etr^{\min}}^{E-E_0} \Gamma(\etr,J) \omega(E-E_0-\etr) d\etr,
\label{eq:kerd}
\end{eqnarray}
in which $\omega$ corresponds to the vibrational density of states of the
product cluster. 

\subsection{Product angular momentum distribution}

We keep the notations introduced previously for the rotational constants
$B_s$, $X$ and $Y$, such that $(B_s,X,Y)=(A,A,B)$ for the prolate top
fragment, and $(B_s,X,Y)=(C,B,C)$ for the oblate top fragment.
We also note $L_X=(\varepsilon^\dagger)^{-1}(\etr-XJ_r^2)$, where
$(\varepsilon^\dagger)^{-1}$ is the reciprocal function of
$\varepsilon^\dagger$.

The probability distribution of $J_r$, $P(J_r;E,J)$ is obtained from
integration of Eq. (\ref{eq:pkjr}) over  $K$ and $\etr$. This quantity
has been investigated in details
in our past work.\cite{pap2} A similar approach can be pursued here,
after using the fact that the RDOS is the sum of the two contributions
$\Gamma_{\rm s}$ and $\pm\Gamma_{\rm t}$. Hence the distribution $P$ can
be written as $P=P_{\rm s}\pm P_{\rm t}$, where the plus (resp. minus)
sign is used for the prolate (resp. oblate) top fragment. The expressions
for the spherical and symmetric top contributions are given by

\begin{eqnarray}
P_{\rm s}(J_r;E,J) &=&\displaystyle
2J_r \int_{\etr^{\min}}^{E-E_0}
\Theta[L_{B_s}(\etr,J_r)-|J-J_r|] \nonumber \\
&& \displaystyle \times \omega(E-E_0-\etr)[ \min(J+J_r,L_{B_s})
-|J-J_r|] d\etr,
\label{eq:psjr}
\end{eqnarray}
and
\begin{eqnarray}
P_{\rm t}(J_r;E,J) &=&\displaystyle  \frac{2}{\sqrt{\lambda}}
\int_{\etr^{\min}}^{E-E_0} \Theta[L_Y(\etr,J_r)-|J-J_r|]
\nonumber \\ &&\displaystyle \times  \omega(E-E_0-\etr) d\etr
\int_{\max(|J-J_r|,L_X)}^{\min(J+J_r,L_Y)} |\erstar-BJ_r^2|^{1/2} dL.
\label{eq:ptjr}
\end{eqnarray}

\subsection{$\cos\theta$ distribution}

The relative orientation of the angular momentum of the product cluster
with respect to its symmetry axis can be calculated by integrating
Eq.~(\ref{eq:pkjr}) over all variables except $K$. At given $\etr$,
$J_r$ and $L$, the number of available $K$ states depends on the values of
$\erstar=\etr-\varepsilon^\dagger(L)$ and $J_r$. In the prolate top
case, there are two states available if $\erstar \geq AJ_r^2$. If
$AJ_r^2\geq \erstar \geq BJ_r^2$, there
are also two states available provided that $|K| \leq K_{\rm max} =
[(\erstar-BJ_r^2)/\lambda ]^{1/2}$, and zero otherwise. The
relative orientation $\cos \theta = |K|/J_r$ must then satisfy
\begin{equation}
\erstar \geq (A\cos^2 \theta + B\sin^2\theta)J_r^2.
\label{eq:ersp}
\end{equation}
The same ideas lead to the following condition for the oblate top
case:
\begin{equation}
\erstar \geq (C\cos^2 \theta + B\sin^2\theta)J_r^2.
\label{eq:erso}
\end{equation}
The integration over the $(L,J_r)$ plane is now restricted, as
indicated in Fig.~\ref{fig:schema}(b), to include the above
conditions. At a given $\cos\theta$, the intersection
$J_r(\theta)$ with the $L=0$ axis is given by $J_r(\theta)=
[\etr/(B\sin^2\theta + B_s\cos^2\theta)]^{1/2}$ in general. This
quantity continuously sweeps the $[J_r(X),J_r(Y)]$ interval.

Finally we get the probability density that the relative orientation
between the angular momentum vector and the revolution axes has the
value $\cos\theta$ within $d\cos\theta$ as
\begin{eqnarray}
P(\cos\theta; E,J) &\propto& \displaystyle
 \int_{\etr^{\min}}^{E-E_0} \omega(E-E_0-\etr) d\etr \times \left\{
\int \!\!\!\!\int_{\erstar \geq XJ_r^2} 2dLdJ_r+ \right. \nonumber \\
& +&  \left. \int\!\!\!\!\int_{XJ_r^2\geq \erstar \geq YJ_r^2}
2\Theta \left[ \erstar - (B\sin^2 \theta + B_s\cos^2\theta)J_r^2 \right]
dL dJ_r \right\},
\label{eq:ct}
\end{eqnarray}
valid for both types of deformations. These distributions, as well
as the preceding probabilities of $\etr$ or $J_r$, must be
calculated numerically in general. Ingredients other than the rotational
densities, most importantly the vibrational densities of states and the
centrifugal energies $\varepsilon^\dagger(L)$, are obtained from the same
techniques described in our previous work.\cite{pap1}

However, the general shape of the distributions can be inferred
already at this stage. In the prolate top situation, $B_s\geq B$ and
$P(\cos\theta)$ is maximal when $\cos\theta=0$, and minimal when
$\cos\theta=1$. The reverse holds for the oblate top case. Therefore,
$P(\cos\theta)$ is a decreasing (resp. increasing) function of
$\cos\theta$ for prolate (resp. oblate) fragments. This means that
the angular momentum gets more likely aligned with the principal axis
with largest inertia. Thus it simply reflects the greater rotational
stability of rigid bodies when their angular velocity is
lower.\cite{goldstein}

\section{Results and discussion}
\label{sec:res}

The clusters we are interested in are Lennard-Jones clusters
characterized by an arbitrary degree of deformation. The simplest
model is that of an ellipsoid, whose axes ratio can take any real
value. At a given (large) number of constituants $n$, the
reasonable assumption of constant density implies that the volume of the
cluster is constant, which allows us to define the equivalent spherical
radius $R_s$ as $R_s = R_0n^{1/3}$, where $R_0$ plays the role of a
lattice parameter. In practice, $R_0$ is taken to reproduce the
rotational constant of the icosahedral 13-atom cluster.

The ellipsoidal shape of the cluster is continuously varied by keeping
its volume constant. By denoting $a$ and $b$ the long parallel and
short perpendicular axes of the ellipsoid, respectively, we quantify
deformation in the cluster using $\gamma = (b-a)/(b+a)$.
Equating $R_s^3$ to $ab^2$ yields
\begin{equation}
\gamma = \frac{1-(a/R_s)^{3/2}}{1+(a/R_s)^{3/2}}.
\end{equation}
As the cluster becomes increasingly prolate, $\gamma$ decreases and tends to
$-1$, which corresponds to a linear system. For oblate systems,
$\gamma$ increases toward $+1$, which corresponds to a planar system.
The rotational constants corresponding to the deformation $\gamma$ are
given by
\begin{equation}
B_s = \frac{5}{4nR_s^2}
\left(\frac{1-\gamma}{1+\gamma}\right)^{2/3}
=B_0 \left(\frac{1-\gamma}{1+\gamma}\right)^{2/3},
\end{equation}
and
\begin{equation}
B = 2B_0 \left[\left(\frac{1-\gamma}{1+\gamma}\right)^{4/3}+
\left(\frac{1+\gamma}{1-\gamma}\right)^{2/3} \right]^{-1}.
\label{eq:y}
\end{equation}

\subsection{LJ$_8$$\longrightarrow$ LJ$_7$+LJ}

The dissociation of LJ$_8$ has been considered first since it provides a
good candidate to test the proposed formalism, the product cluster LJ$_7$
being not perfectly spherical in its ground state geometry. For this
system, the anharmonic VDOS and the effective dissociation potential
were obtained from parallel tempering Monte Carlo and Wang-Landau
simulations, respectively. The technical details can be found in our
previous work.\cite{pap1}

\begin{figure}[htb]
\vbox to 5cm{
\includegraphics{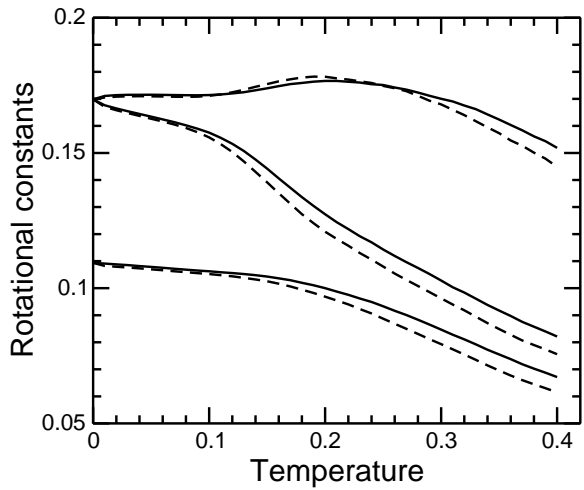}
\vfill}
\caption{Thermally averaged rotational
constants of the LJ$_7$ cluster versus temperature, for
initially nonrotating (solid lines) and rotating ($J=3$, dashed
lines) systems.}
\label{fig:B_T}
\end{figure}

To quantify the real extent of deformation in the vibrationally excited
LJ$_7$ cluster, we have analysed the thermal evolution of its
rotational constants. In Fig~\ref{fig:B_T} the average constants
are plotted versus $T$. At $T$=0, the minimum energy configuration is
the pentagonal bipyramid, an exactly oblate system for which $A=B
\approx 0.17$, $C\approx 0.11$. These rotational constants yield a
deformation index $\gamma\approx 0.3$, or $a/R_s=0.7$ with $R_s=0.55$.
At higher temperatures, the cluster deviates more and more from the
oblate shape, eventually becoming prolate. In order to get meaningful
statistics of the dissociation process using standard molecular
dynamics, we had
to thermalize the parent cluster at rather high temperatures,
typically above the melting point ($T\approx 0.2$). The rotational
constants used in the PST analysis were taken at this precise
temperature, yielding $A\approx 0.18$ and $B=C\approx 0.11$ LJ units.
These values give an equivalent deformation index $\gamma=-0.3$ in the
ellipsoidal picture, or $a/R_s=1.5$. Interestingly, the double
icosahedron LJ$_{19}$ is more spherical than LJ$_7$ ($\gamma\approx
-0.18$), and remains so even above its melting temperature.

In Table~\ref{tab:LJ7} we have reported the values of $\langle\etr
\rangle$ and $\langle J_r\rangle$ obtained by MD simulations and by
the PST descriptions in various rigid body approximations for LJ$_7$.
These results
correspond to $E/n=1.2$ and $J=0$, 1, 2 and 3 (one LJ unit of angular
momentum approximately equals 33$\hbar$ for argon). The data for $J=0$ is
obtained using simplifications to Eqn.~(\ref{eq:kerd}--\ref{eq:ct}),
which arise due to the new restraint on the orbital momentum,
namely $L=-J_r$.\cite{klots,cb,pap2,pap3}

Under these aforementioned conditions, increasing cluster
deformation ($\gamma<0$ or $\gamma>0$) mainly
increases the product angular momentum, and marginally reduces the
KER. The prolate top approximation performs significantly better than
the simple spherical assumption in reproducing the MD results, but
this is also true for the oblate top approximation.

\begin{figure}[htb]
\vbox to 8cm{
\includegraphics{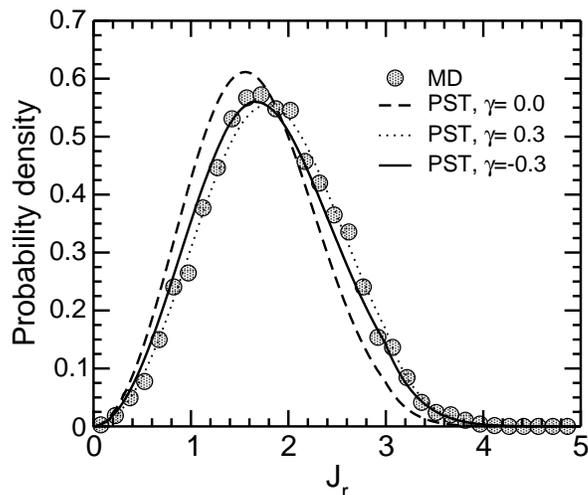}
\vfill}
\caption{Distribution of final angular momentum after dissociation of
LJ$_8$, from different approximations (sphere-atom, prolate-atom and
oblate-atom) for the rotational DOS in the PST calculation.}
\label{fig:distriJ_7}
\end{figure}

The complete distribution of $J_r$ contains more information than 
the average value alone. This distribution is plotted in
Fig.~\ref{fig:distriJ_7} for the initial angular momentum $J=3$.
A good agreement between MD and PST is obtained when the deformation
of the product cluster is taken into account. By comparison,
the distribution in the spherical approximation underestimates $J_r$.
More interestingly, the PST/oblate calculation looks very close
to the prolate result, suggesting that the type of deformation does not
alter significantly the rotational distribution after evaporation.
The KER distribution, not plotted here, shows a similar behavior.
However, it should be kept in mind here that we are dealing with
only moderate deviations from sphericity.

\begin{figure}[htb]
\vbox to 8cm{
\includegraphics{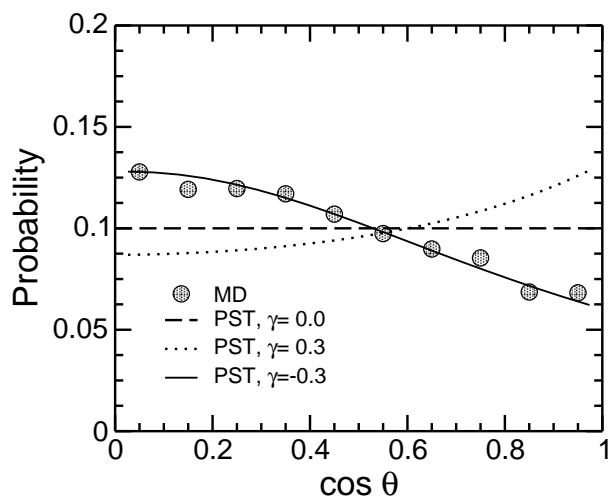}
\vfill}
\caption{Distribution of $\cos\theta$ in the dissociation of LJ$_8$
at $J$=3 and $E/n$=1.2. The results of MD simulations are compared to
PST calculations for different approximations of the LJ$_7$ main product.}
\label{fig:districostheta_7}
\end{figure}

More insight is gained from the distribution of $\cos\theta$,
represented in Fig.~\ref{fig:districostheta_7} for the same conditions
of total energy and angular momentum. The distribution is obviously
uniform in the spherical case, as there is no privileged axis.
As expected from our previous analysis, the prolate description of
the product cluster favors low $\cos\theta$. This is in agreement
with MD, but differs from the PST result in the oblate approximation.

\begin{figure}[htb]
\vbox to 8cm{
\includegraphics{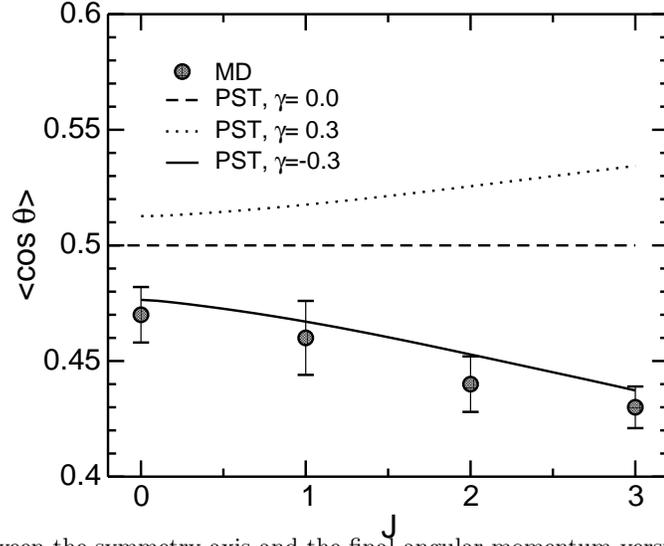}
\vfill}
\caption{Average angle $\langle \cos\theta\rangle$ between the symmetry
axis and the final angular momentum versus initial $J$ in the
dissociation of LJ$_8$ with $E/n$=1.2.}
\label{fig:costhetamoyen_7}
\end{figure}

Increasing angular momentum in the parent cluster brings the final
rotation axis closer to the major principal axis, as shown in
Fig.~\ref{fig:costhetamoyen_7}. This emphasizes the need for properly
describing deformations in clusters having a significant or even
moderate rotational motion. The values of $J$ considered in this
figure are small enough to produce rotational
heating.\cite{pap2,stace} However high angular momenta, which result
in rotational cooling, also display a similar behavior. Comparing the
values of $\langle\cos\theta\rangle$ at $J=0$ and $J=3$ shows that the
deviation from random orientation (for which $\langle\cos\theta
\rangle=1/2$) is half due to the extent of deformation, and half due
to angular momentum itself. From this figure, but also from
Fig.~\ref{fig:districostheta_7}, alignment between the rotation axis
and the main symmetry axis is about twice more efficient for prolate
tops than for oblate tops. Such an effect is expected to be also
magnified by the extent of deformation.

\subsection{Size effects in model clusters}

We now discuss the general effects of deformation on the
statistical properties of larger clusters after evaporation. Our
interest here is {\em not}\/ to reproduce MD or experimental results,
for this reason we have used an approximate model, where the $n$-atom
LJ cluster is treated as a continuous medium with constant density, and
the vibrational density of states is assumed to be harmonic. Such a
model allows us to vary arbitrarily the degree of deformation
$\gamma$ in the whole range $-1< \gamma< 1$. The interaction
potential is also simplified as $-4n/r^6$, which yields a barrier
energy $\varepsilon^\dagger$ proportional to $L^3$. As deformation
of the cluster increases, the physical extent of the cluster also
increases and we expect the $1/r^6$ approximation to fail eventually.
A more appropriate potential such as $1/(r-r_0)^6$ would not
be relevant for the present, qualitative discussion, even though it would
provide a more robust ground for eventual comparisons.

\begin{figure}[htb]
\vbox to 8cm{
\includegraphics{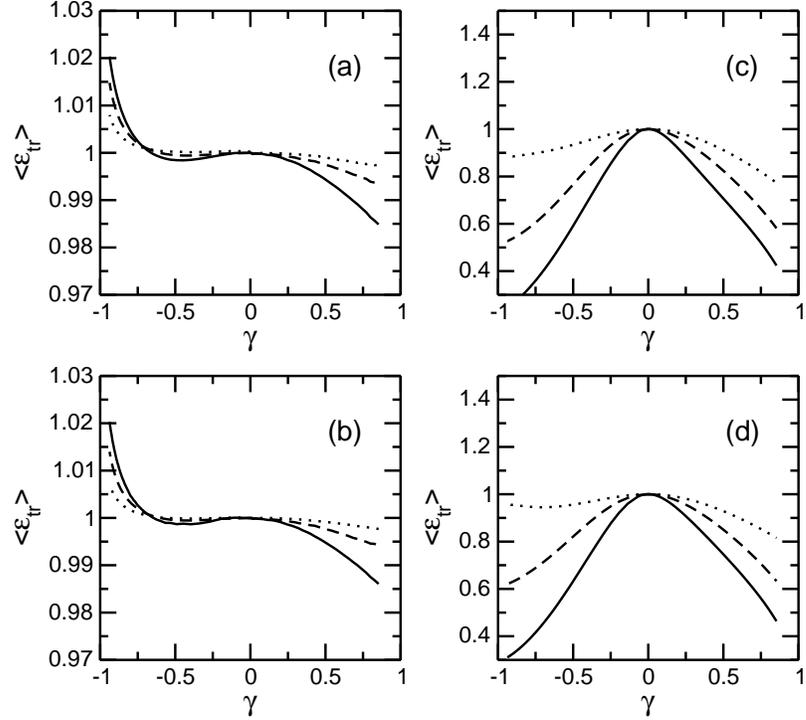}
\vfill}
\caption{Average KER $\langle\etr\rangle$ versus deformation $\gamma$
in the evaporation of model LJ$_{n+1}$ clusters under several conditions.
(a) $E/n=1.2$ and $J=0$; (b) $E/n=1.2$ and $J=20$; (c) $E/n=0.9$ and
$J=0$; and (d) $E/n=0.9$ and $J=20$. The results are shown for $n=50$
(solid lines), $n=100$ (dashed lines), and $n=200$ (dotted lines).}
\label{fig:etr}
\end{figure}

\begin{figure}[htb]
\vbox to 9cm{
\includegraphics{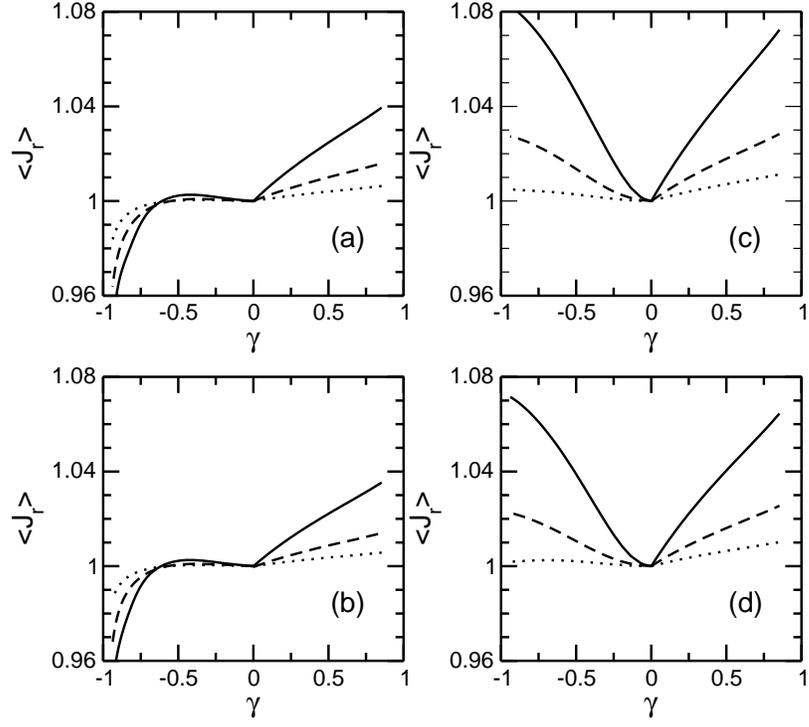}
\vfill}
\caption{Average product angular momentum $\langle J_r\rangle$ versus
deformation $\gamma$ in the evaporation of model LJ$_{n+1}$ clusters
under the same conditions as in Fig.~\protect\ref{fig:etr}. The same
graphs conventions are also used.}
\label{fig:Jr}
\end{figure}

Three parent cluster sizes $n+1$ have been selected, which correspond
to $n=50$, 100, and 200, respectively.
To cover a vast range of situations, two total energies ($E/n=0.9$ and
1.2 LJ units per atom) and two angular momenta ($J=0$ and $J=20$ LJ
units) have been considered, thus providing four different physical
conditions for fragmentation. The effects of deformation on the
kinetic energy released and on the final angular momentum are shown
in Figs.~\ref{fig:etr} and \ref{fig:Jr}, respectively, for the four
present situations. From a general point of view, smaller clusters
show more pronounced deviations of the average KER, product
angular momentum. This is easily understood by
noticing that the rotational constants, hence all the parameters
which characterize the deformation such as $\lambda$, decrease
with size very fast ($\propto n^{-5/3}$) with respect to, e.g.
the centrifugal barrier energy ($\propto n^{-1/2}$). In other terms,
the specific symmetric top contributions $\gamma_{\rm t}$ to the RDOS
become less important with increasing size.

It is useful to analyse first the $J=0$ case, since all
the integrals involved are one-dimensional. For convenience, we assume
in the following discussion that the main product is a prolate top.
In this case the RDOS is exactly given for $J=0$ and a $C_6/r^6$
interaction potential by\cite{pap2}
\begin{eqnarray}
\Gamma(\etr,J=0) &=& \int_0^{J_r^A} 2J_r dJ_r \nonumber \\
&+& \int_{J_r^A}^{J_r^B} 2\left[ \frac{\etr - aJ_r^3 -
BJ_r^2}{\lambda}\right]^{1/2} dJ_r,
\label{eq:pe}
\end{eqnarray}
where $aJ_r^3 = \varepsilon^\dagger(J_r)$ is the centrifugal energy at
the barrier. Here $a$ does not depend on $\gamma$ but only on $n$ and
$C_6$. $J_r^A$ and $J_r^B$ are solution of $\etr = A J_r^2 + aJ_r^3$
and $\etr = BJ_r^2 + aJ_r^3$, respectively. When $a=0$, this integral
can be exactly solved as
\begin{equation}
\Gamma(\etr,J=0) = \frac{\etr}{\sqrt{B\lambda}}\left[ \frac{\pi}{2}
-\sin^{-1}\sqrt{\frac{B}{A}}\right].
\label{eq:gammaazero}
\end{equation}
$\Gamma$ is a linear function of $\etr$ independently of the
rotational constants, therefore the average KER does not change with
deformation, because of the normalization of the integrals involved in
the average. This shows that any variation in the average energy
released is due to the centrifugal barrier only. The Klots
models\cite{klots} approximate the rotational densities
$\Gamma(\etr,{\rm low~}J)$ as a power law in $\etr$, the
exponent increasing proportionally with the number of rotational
degrees of freedom. Since this number does not depend on the possible
symmetry of the products, there are no deformation effects on the
evaporation statistics within the Klots model.

We have not found simple expressions for the RDOS by Taylor expanding
the general formula of Eq.~(\ref{eq:pe}) in $a$, because the result
still contained complex functions of the rotational constants.
Instead we consider the $J_r$ observable. The general
expression for the probability distribution of $J_r$ at $J=0$, for
prolate top product, reads\cite{pap2}
\begin{eqnarray}
P(J_r;J=0) &=& 2J_r\int_{AJ_r^2+aJ_r^3}^{E-E_0} \omega(E-E_0-\etr)
d\etr \nonumber \\ &+& \int_{BJ_r^2+aJ_r^3}^{AJ_r^2+aJ_r^3}
2\left[ \frac{\etr - aJ_r^3 - BJ_r^2}{\lambda}\right]^{1/2}
\omega(E-E_0-\etr) d\etr.
\label{eq:gammaj0}
\end{eqnarray}
After some algebra, the lowest order in expansion in $|\gamma|^{1/2}$
of this expression is found to be
\begin{eqnarray}
P(J_r;J=0) &\simeq& 2J_r \int_{B_0 J_r^2 + aJ_r^3}^{E-E_0} \omega(E-E_0
-\etr)d\etr \nonumber \\
&+& 2\sqrt{2}B_0^{3/2} |\gamma|^{3/2} J_r^4 \omega'(E-E_0-B_0J_r^2-aJ_r^3),
\label{eq:pejr}
\end{eqnarray}
where $\omega'=d\omega/dE$. As $|\gamma|$ increases, the weight of the
$J_r^4$ term increases as well as $\langle J_r\rangle$. However,
because the correction has the power 3/2, the deviation with respect
to the sphere remains small.

A similar treatment can also be carried ou in the case of oblate top
products. However, the result differs because the smaller rotational constants
changes from $B \simeq B_0(1+4\gamma/3)$ for $\gamma<0$ to
$B_s \simeq B_0(1-2\gamma/3)$ for $\gamma>0$:

\begin{eqnarray}
P(J_r;J=0) &\simeq& 2J_r \int_{B_0 J_r^2 + aJ_r^3}^{E-E_0} \omega(E-E_0
-\etr)d\etr \nonumber \\
&+& \frac{4B_0}{3} \gamma J_r^3 \omega(E-E_0-B_0J_r^2-aJ_r^3),
\label{eq:oejr}
\end{eqnarray}
The corrective term is still positive, and favors higher values of
$J_r$ at increasing $\gamma$. However, this term grows linearly with
$\gamma$, therefore the correction is more important for oblate tops
than for prolate tops.

The increase in product angular momentum, as well as the non
equivalence of the two deformations close to the spherical shape,
are indeed seen in Figs.~\ref{fig:Jr}(a,b), and these effects are also
observed at nonzero initial angular momenta, Figs.~\ref{fig:Jr}(c,d).
Obviously, other parameters to quantify the extent of deformation
might affect the shape of the curves $\langle J_r\rangle(\gamma)$ or
$\langle \etr\rangle(\gamma)$. For instance, if $\gamma$ was defined from
the inertia momenta or rotational constants directly, $\langle
J_r\rangle$ would vary differently around zero, because the inertia
momenta vary quadratically with the axes lengths $a$ and $b$.

Increasing $|\gamma|$ tends to favor larger product angular momenta,
which is a consequence of smaller rotational constants, the same
rotational energy being achieved through higher $J_r$. In turn, the
centrifugal barrier increases and the kinetic energy released
decreases. This explains the correlation between $\langle\etr\rangle$
and $\langle J_r\rangle$ seen in Figs.~\ref{fig:etr} and \ref{fig:Jr}.
Larger total energies decrease the relative importance of the
rotational contribution, as well as the effects of nonsphericity. This is
precisely what we observe in Figs.~\ref{fig:etr}(b) and \ref{fig:Jr}(b).
Conversely, if now we increase $J$ keeping the total energy fixed,
more kinetic energy is allowed in the rotational modes, which usually
yields amplified deviations with respect to the spherical reference
for both the KER and the product angular momentum.

At finite angular momentum $J$, it is much harder to provide simple
explanations for the behavior of the statistical quantities, because
new parameters come to play an important role. Increasing $J$
generally has a much more dramatic effect than increasing $E$.
The minimum value for the KER, $\etr^{\rm min}$, has a component which
grows linearly with the smallest rotational constant of the
product.\cite{cb,pap2} At large deformations, either prolate or
oblate, $\etr^{\rm min}$ always takes values smaller than for the
spherical reference. Therefore low values of the KER are further
statistically favored. This effect may be larger than the small
increase due to the centrifugal energies seen in
Fig.~\ref{fig:etr}(a).
Second, because high deformations are characterized with at least
one very small rotational constant, the upper limits $J_r(Y)$ of the
product angular momentum in the $(L,J_r)$ integration range can reach
higher values. Thus the integration range extends to larger $J_r$, and
the average value of $\langle J_r\rangle$ increases.

The competition between these various effects, as well as the
different functional forms for the RDOS, provide a rather rich
variety of behaviors for $\langle\etr\rangle$ and $\langle
J_r\rangle$, as shown in Figs.~\ref{fig:etr} and \ref{fig:Jr}.
The exact numerical calculations displayed in
Figs.~\ref{fig:etr}(c,d) and \ref{fig:Jr}(c,d)
confirm that much larger deviations are seen with respect to the
spherical reference when $J$ is nonzero. In particular, the marked
decreases in the KER and the correlated increases in the
product angular momentum are of larger magnitude than for $J=0$. 
A notable result is that the most important differences
between the two types of deformation (oblate and prolate) are
exclusively seen at very low total momentum and for large sizes.
These differences are related to variations in small
rotational constants. Above some specific total energy and/or some total
angular momentum, the effects of prolate or oblate deformations become
rather similar in shape and magnitude, except very close to $\gamma=0$.

\begin{figure}[htb]
\vbox to 8cm{
\includegraphics{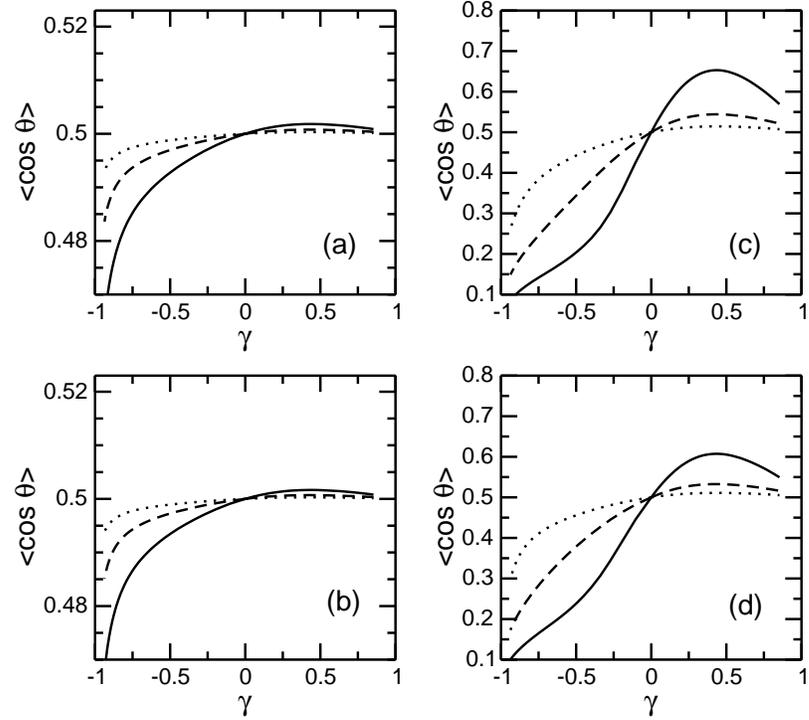}
\vfill}
\caption{Average orientation $\langle \cos\theta \rangle$ between the
symmetry axis and the product angular momentum vector versus
deformation $\gamma$ in the evaporation of model LJ$_{n+1}$ clusters
under the same conditions as in Fig.~\protect\ref{fig:etr}. The same
graphs conventions are also used.}
\label{fig:costheta}
\end{figure}

Another interesting feature of these nonspherical systems is their
ability to adopt a preferential direction for the rotational
excitation following evaporation, as measured with respect to the
cylindrical axis. The average cosine of the angle between
the product angular momentum and the symmetry axis, as obtained from
Eq.~(\ref{eq:ct}), is shown in
Fig.~\ref{fig:costheta} for the same mechanical conditions as in
Figs.~\ref{fig:etr} and \ref{fig:Jr} versus the extent of deformation
$\gamma$. Again, the effects monotonically decrease with increasing
cluster size.

From the previous discussion we already know
that $\langle\cos\theta\rangle$ is positive (resp. negative) for
oblate (resp. prolate) deformations. Hence, the variations of
this quantity seen in Fig.~\ref{fig:costheta} are not surprising, at
least for small deformations. A simple, first-order perturbative
expansion in the equation (\ref{eq:ct}) can
be used as a check of the behavior of $\langle\cos\theta\rangle$ close
to $\gamma\sim 0$. However, for arbitrary total energies and angular
momenta, the extent of alignment $\langle\cos\theta\rangle$ always
shows the same qualitative variations with respect to $\gamma$.
For oblate top products, dissociation induces rotation around a
long axis preferentially, and the effect is more significant at
smaller total energies or larger initial angular momenta.
Prolate top products, on the other hand, display a non-monotonic
behavior, concomitant with the variations of the large
rotational constant, Eq.~(\ref{eq:y}), which has a maximum when
$\gamma=(\sqrt{2}-1)/(\sqrt{2}+1)$. The fact that alignment is
hindered at very large oblate deformations is a simple consequence
that the two rotational constants $B$ and $C$ tend to 0, hence
the second integral over $L$ and $J_r$ in Eq.~(\ref{eq:ct}) goes
to 0, and the system becomes equivalent to a sphere again from the
point of view of its orientation. Therefore, it is practically
impossible to reach a perfect statistical alignment for oblately
deformed systems. Such a situation is never met with
prolate deformed clusters, since one rotational constant always
keeps increasing for large deformations. The shape of
the curves in Figs.~\ref{fig:costheta}(c,d) also reveals that
alignment can be quite effective, even at moderate deformations.
For a 50-atom cluster, $J\sim 20$ is close to the typical thermal
angular momentum near the melting point, and a deformation of about
20\% is enough to induce statistical alignment beyond 25\%.

Finally, following our previous work,\cite{pap2} we focus on the
influence of nonsphericity on the rotational cooling and heating
effects. The same simple model for LJ clusters was used in the
harmonic approximation for the VDOS and the $C/r^6$ radial
potential for the dissociation energy. We consider here
more realistic conditions for dissociation, namely thermal
distributions for the energy and angular momentum of the parent
cluster, as well as a possible extra energy shift to model an initial,
brief excitation (from photoabsorption or collision). To provide
ground for comparison, these distributions are assumed to be
Boltzmann-like and identical for all geometries. The final
distributions were calculated
by taking a full account of the prolate or oblate character of the
main product cluster.

\begin{figure}[htb]
\vbox to 11cm{
\includegraphics{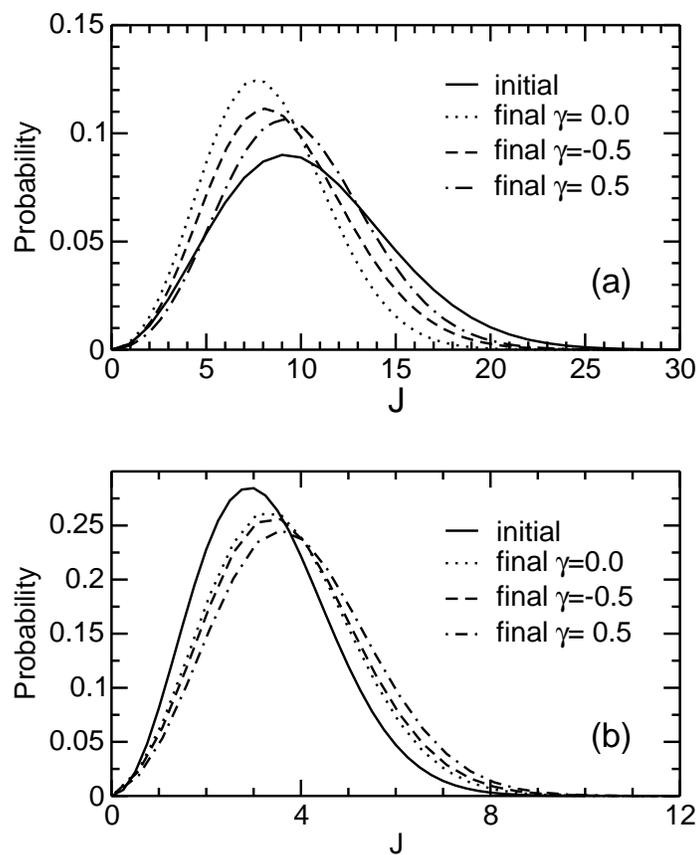}
\vfill}
\caption{Initial and final distributions of angular momentum after
evaporation from a 51-atom model LJ cluster, the main product being
assumed spherical $(\gamma=0.0)$, prolate $(\gamma=-0.5)$, or oblate
$(\gamma=0.5)$. (a) Hot thermal distributions of energy and angular
momentum $(T=0.5)$; (b) cold thermal distributions $(T=0.05)$ and
extra vibrational excitation ($\Delta E=30$).}
\label{fig:rot}
\end{figure}

In spherical clusters, we have seen that a purely thermal
distribution generally leads to rotational cooling, {\em i.e.} the
distribution of angular momentum of the product cluster is shifted
to lower values.\cite{pap2} On the
other hand, an initially cold cluster submitted to a sudden vibrational
excitation preferentially exhibits rotational heating.\cite{pap2,stace}
In Fig.~\ref{fig:rot} the final angular momentum distributions are
compared for the three types of products. Even though the product
clusters are significantly deformed, the general rotational
behavior remains similar to that of the spherical top. As seen from
Fig.~\ref{fig:rot}(a), purely thermal evaporation induces rotational
cooling, and this effect is attenuated for nonspherical clusters.
Deformation also influences the final distribution itself, which differs
significantly from the Boltzmann behavior for prolate systems. These
results are consistent with our previous analysis, as thermal
evaporation corresponds more likely to the cases (c) and (d) of
Figs.~\ref{fig:etr}--\ref{fig:costheta} with substantial initial
$J$. At $|\gamma|=0.5$, Fig.~\ref{fig:Jr}(d) shows that $\langle
J_r\rangle$ increases by about 5\% for oblate top products, slightly
less for prolate top products, with respect to the spherical
case. These changes (including their magnitude) are reflected on the
distributions in Fig.~\ref{fig:rot}(a).

Rotational heating is produced in a different way, by
adding a sharp extra vibrational excitation to an initially cold
cluster. Deformed clusters also exhibit rotational heating, as shown
in Fig.~\ref{fig:rot}(b). For both prolate and oblate top products,
deformation further heats the final rotational motion, but the two
deformations behave dissimilarly, the effect being much stronger for the
oblate top product. This is understood by considering again
Fig.~\ref{fig:Jr}(b), which shares the mechanical conditions of high
total energies but low angular momenta. The increase in $\langle
J_r\rangle$ is quite important for oblate tops, although it is barely
noticeable for prolate tops at $|\gamma|=0.5$.

Deformation most often induces higher rotational excitations
upon evaporation, this explains why it reduces rotational cooling, but
enhances rotational heating.

\section{Conclusion}
\label{sec:ccl}

The present work was aimed at providing a quantitative study of the
effects of cluster shape on its evaporation statistics. Within the
framework of phase space theory, the distributions of kinetic energy
released and final angular momenta were obtained assuming the fragments
could be described as an atom plus a prolate or an oblate top as the
main product.

These theoretical tools were challenged on the example of the
unimolecular dissociation of the 8-atom Lennard-Jones cluster, chosen
for its significantly oblate shape. The small disagreement reported
earlier\cite{pap1} between PST and molecular dynamics simulations for
the distribution of final angular momentum was solved after accounting
for the nonspherical character of LJ$_7$.

Using simple models for the vibrational density of states and the
dissociation potential, we systematically investigated the changes
induced by deformation. We found very significant deviations with
respect to
the spherical case for both the final angular momentum and kinetic
energy, especially for nonzero initial momenta. In such cases, the
relative variations of the energetic properties are larger than those
of angular momenta by up to even one order of magnitude.

We could interpret these changes as originating partly from the
specific centrifugal barrier in our present implementation of PST, but
also from the new energy constraints arising with lower rotational
constants. These effects are not included in simpler statistical rate
theories of unimolecular dissociation such as the Klots models, not
mentioning the Engelking-Weisskopf\cite{engelking,weisskopf} or RRK
approaches.\cite{rrk}

We also obtained information about the relative orientation of the
product angular momentum, by considering restricted integration of the
rotational densities of states. Alignment toward the long axes is
generally observed for all deformations, in agreement with mechanical
arguments. This alignment was seen to be very effficient already at
moderate initial angular momenta, and for moderate deformations.

Finally, we discussed the influence of deformation on the rotational
cooling and heating effects resulting from specific excitations. While
cooling is attenuated for nonspherical clusters, heating is
amplified. These behaviors are attributed to the decrease of one
rotational constant, which favors larger angular momenta in the
product cluster.

The examples considered here were obtained on model clusters, but we
expect the present results to be relevant for more realistic
systems. Charged rare-gas clusters, which display very prolate shapes
below 15 atoms due to the presence of a linear ionic trimer, have been
the subject of intense experimental and theoretical activity from the
point of view of their dissociation patterns.\cite{rg}

Polycyclic aromatic hydrocarbon (PAH) molecules are strongly oblate
and their rotation was considered within the astrophysical context
by Rouan and co-workers.\cite{rotpah} Dehydrogenation of these
molecules, in particular, has received a special attention among
several experimental and theoretical groups.\cite{ho,boissel,dibben}
This problem could be investigated with the same methods used here.

In these two examples, a realistic atomistic description requires
models much more involved than simple pairwise potentials. This practical
limitation prevents one fully relying on brute-force simulations, thus making
statistical approaches most useful in the future.

\section*{Acknowledgments}

The authors wish to thank the GDR {\it Agr\'egats, Dynamique et
R\'eactivit\'e} for financial support.

\clearpage
\begin{table}
\begin{tabular}{|c|c|c|c|c|c|c|c|c|}
\colrule
& \multicolumn{2}{c|}{spherical} & \multicolumn{2}{c|}{oblate} &
\multicolumn{2}{c|}{prolate} &  \multicolumn{2}{c|}{MD} \\
\colrule
& $\langle\etr\rangle$ & $\langle J_r\rangle$ & $\langle\etr\rangle$
& $\langle J_r\rangle$ & $\langle\etr\rangle$ & $\langle J_r\rangle$
& $\langle\etr\rangle$ & $\langle J_r\rangle$ \\
\colrule
$J=0$ & 0.51 & 1.07 & 0.51 & 1.16 & 0.51 & 1.12 & 0.50 & 1.11 \\
\colrule
$J=1$ & 0.56 & 1.21 & 0.56 & 1.31 & 0.56 & 1.27 & 0.55 & 1.25 \\
\colrule
$J=2$ & 0.70 & 1.45 & 0.70 & 1.58 & 0.69 & 1.52 & 0.68 & 1.52 \\
\colrule
$J=3$ & 0.89 & 1.65 & 0.89 & 1.85 & 0.87 & 1.80 & 0.89 & 1.84 \\
\colrule
\end{tabular}
\caption{Average kinetic energy released $\langle\etr\rangle$ and
product angular momentum $\langle J_r\rangle$ after evaporation of the
LJ$_8$ cluster at total excitation energy $E/n=1.2$ and several total
angular momenta. The values obtained from molecular dynamics
simulations are compared to the PST predictions in the spherical
($\gamma=0$), prolate top ($\gamma=-0.3$) and oblate top ($\gamma=0.3$)
approximations for the product cluster LJ$_7$.}
\label{tab:LJ7}
\end{table}

\clearpage

\vfill

\clearpage

\vfill

\clearpage

\vfill

\clearpage

\vfill

\clearpage

\end{document}